\documentclass[9pt,twocolumn,twoside]{osajnl}

\journal{jocn} 

\setboolean{shortarticle}{false}

\title{QoT Estimation using EGN-assisted Machine Learning for Multi-Period Network Planning}

\author[1,2,*]{Jasper Müller}
\author[1,2]{Sai Kireet Patri}
\author[1]{Tobias Fehenberger}
\author[1]{Helmut Griesser}
\author[1]{Jörg-Peter Elbers}
\author[2]{Carmen Mas-Machuca}

\affil[1]{ADVA, Fraunhoferstr. 9a, 82152 Martinsried/Munich, Germany}
\affil[2]{Chair of Communication Networks, Technical University of Munich, Arcisstr. 21, Munich, Germany}

\affil[*]{Corresponding author: jmueller@adva.com}

\usepackage{tikz}
\newcommand\copyrighttext{%
  \footnotesize \textcopyright~2022~Optica Publishing Group. One print or electronic copy may be made for personal use only. Systematic reproduction and distribution, duplication of any material in this paper for a fee or for commercial purposes, or modifications of the content of this paper are prohibited.
  \href{https://doi.org/10.1364/JOCN.472632}{DOI: 10.1364/JOCN.472632}.
  }
\newcommand\copyrightnotice{%
\begin{tikzpicture}[remember picture,overlay]
\node[anchor=north,yshift=-40pt] at (current page.north) {\fbox{\parbox{\dimexpr\textwidth-\fboxsep-\fboxrule\relax}{\copyrighttext}}};
\end{tikzpicture}
\vspace{-0.3cm}
}

\usepackage{color}
\usepackage{xcolor}
\usepackage{graphicx}
\usepackage{tikzscale}
\usepackage{tikz}
\usepackage{pgfplots}
\usepackage{caption}
\usepackage{subcaption}
\usepackage{verbatim}
\usepackage{multirow}

\usetikzlibrary{plotmarks,matrix,chains,scopes,fit,calc,shapes,positioning,decorations,intersections,fit,backgrounds,patterns}
\usetikzlibrary{fadings,shapes.arrows,shadows}
\pgfplotsset{compat=newest} 
\usepackage{forest}
\usepgfplotslibrary{groupplots}

\pgfplotsset{plot coordinates/math parser=false}
\pgfplotsset{every axis plot/.append style={solid,line width=1.5pt,mark size=1.5pt,mark options={solid,fill=white}}}
\pgfplotsset{every axis legend/.append style={legend cell align=left,font=\footnotesize}}
\newlength\FigureWidth
\newlength\FigureHeight
\newlength\FullFigureWidth
\graphicspath{{figures/}}

\pgfplotsset{myLegend/.append style={legend style={font=\footnotesize,at={(0.5,0.98)},anchor=north,align=left,legend columns=3}}}

\usepackage{xparse}
\tikzfading[name=arrowfading, top color=transparent!0, bottom color=transparent!95]
\tikzset{arrowfill/.style={#1,general shadow={fill=black, shadow yshift=-0.8ex, path fading=arrowfading}}}
\tikzset{arrowstyle/.style n args={3}{draw=#2,arrowfill={#3}, single arrow,minimum height=#1, single arrow,
single arrow head extend=.3cm,}}

\NewDocumentCommand{\tikzfancyarrow}{O{2cm} O{FireBrick} O{top color=OrangeRed!20, bottom color=Red} m}{
\tikz[baseline=-0.5ex]\node [arrowstyle={#1}{#2}{#3}] {#4};
} 

\makeatletter
\tikzset{
    block filldraw/.style={
        draw, fill=yellow!20},
    block rect/.style={
        block filldraw, rectangle},
    block/.style={
        block rect, minimum height=0.8cm, minimum width=6em},
    from/.style args={#1 to #2}{
        above right={0cm of #1},
        /utils/exec=\pgfpointdiff
            {\tikz@scan@one@point\pgfutil@firstofone(#1)\relax}
            {\tikz@scan@one@point\pgfutil@firstofone(#2)\relax},
        minimum width/.expanded=\the\pgf@x,
        minimum height/.expanded=\the\pgf@y}}
\makeatother

\newcommand{\Ptx}{\ensuremath{P_{\text{tx}}}}
\newcommand{\SNR}{\ensuremath{\text{SNR}}}

\newcommand{\rev}[1]{{#1}}
\newcommand{\oldrev}[1]{{#1}}
\newcommand{\oldrevv}[1]{{#1}}
\dates{Accepted Nov 7, 2022}

\doi{\url{http://dx.doi.org/10.1364/JOCN.472632}}

\begin{abstract}
\oldrev{The rapidly growing traffic demands in fiber-optical networks require flexibility and accuracy in configuring the lightpaths, for which fast and accurate quality of transmission (QoT) estimation is of pivotal importance. This paper introduces a machine learning (ML) based QoT estimation approach that meets these requirements. The proposed gradient-boosting ML model uses precomputed per-channel self-channel-interference values \oldrevv{as representative and condensed features to estimate non-linear interference in a flexible-grid network}. With enhanced GN model simulation as baseline, the ML model achieves a mean absolute signal-to-noise ratio (SNR) error of approximately 0.1~dB, which is an improvement over the GN model. For three different network topologies and network planning approaches of varying complexity, a multi-period network planning study is performed in which ML and GN are compared as path computation element (PCE). The results show that the ML PCE is capable of matching or slightly improving the performance of the GN PCE on all topologies \rev{while reducing significantly the computation time of the network planning by up to 70\%.}} 
\end{abstract}

\setboolean{displaycopyright}{true}

\begin{document}

\maketitle
\copyrightnotice

\section{Introduction}
In recent years, the demand in capacity for optical networks has grown rapidly \cite{cisco}. In addition, the commercial deployment of flexible grid Bandwidth Variable Transceivers (BVTs) greatly increased the amount of possible network configurations. The optimization of the current network resources and the reduction of conservative margins becomes increasingly important as it is a cost-efficient solution to meet the rising internet traffic \cite{acacia}.

The optimal use of the current network resources aims at placing and configuring lightpaths to maximize the network capacity while meeting the traffic demands.
Finding the optimal solution using tools like integer linear programming (ILP) is not feasible for complex network topologies due to the high computational complexity. Therefore routing, configuration and spectrum assignment~(RCSA) algorithms are used in physical-layer aware network planning for optimized placement and configuration of lightpaths. The RCSA relies on a path computation element (PCE) to estimate the quality of transmission (QoT) of a lightpath and therefore determine its validity \cite{patri2020planning}. Therefore the accuracy of the QoT estimation directly \oldrev{affects} the needed margins to safely ensure error free transmission. The computation time of the QoT estimate affects the number of configurations that can be investigated during planning and therefore the degree of optimization that can be achieved. 

A crucial part for QoT estimation is the non-linear interference (NLI) computation. A variety of physical models are available, such as the split-step simulations and various Gaussian noise (GN) models. These follow a trade-off between accuracy and computational complexity with split-step simulations being the most accurate and complex model and the closed-form GN model \cite{GGN} the least accurate but fastest model. In a C-band system using high dispersion fibers such as standard single mode fiber, the main contributions to NLI are self-channel interference (SCI) and cross-channel interference (XCI). The majority of the computational complexity of accurate models like the enhanced GN (EGN) model \cite{egn} comes from the XCI computation.

Machine learning (ML) has been extensively investigated as a tool for QoT estimation. ML is known for its ability to accurately learn non-linear relations and its fast computation after the initial training. Therefore ML fits well to the requirements on a QoT estimator that is used for network planning applications. \oldrev{A general introduction and overview to ML applications in optical communications and networking is provided \cite{KHAN}.} Different approaches to ML based QoT estimation have been investigated. They have been reviewed and compared in a survey \cite{Pointurier}. 
Most ML approaches in the literature fall into the category of supervised learning. Both classification and regression approaches have been investigated. Classification methods, such as support vector machines, neural networks \oldrev{(NNs)} and random-forest algorithms, have shown high accuracy in predicting the feasibility of a candidate lightpath in relation to a BER threshold \cite{Aladin, Rottondi, Morais, Safari:21}. 
Regression approaches, on the other hand, are allowing for a quantitative QoT estimate. In this context, the estimation of a generalized signal-to-noise ratio (GSNR) \cite{Usmani} as well as GSNR distributions \cite{Ibrahimi} for assumed imperfect representation of physical parameters by the ML features, have been investigated. For multiple channels on a testbed link, Q-factor prediction has been demonstrated \cite{Gao}. Regression has also been used for the modeling of parameter uncertainty \cite{Pesic}. \oldrev{In \cite{Zhuge} a hybrid approach is chosen. The output of a GN model and physical modeling and monitoring parameters are used as input for an NN. The approach is shown to reduce estimation errors due to the inaccuracy of the GN model as well as errors due to introduced uncertainty in the launch power.} Furthermore, an ML-based QoT estimation regression model has been compared to the closed-form GN model in the context of network planning \cite{Lonardi}. The closed-form GN model was considered as ground truth and parameter uncertainty is added, with the level of parameter uncertainty in a network assumed to be known. \oldrevv{The authors have shown that accuracy is not sufficient as the sole evaluation metric to determine the performance of a QoT estimator. It is therefore important to test the model in the context of network planning.} To ensure extensive validation, the ML model was tested in a network planning task and full error distributions were considered when evaluating its accuracy on the simulation data.
A neural-network-based NLI regression model has been demonstrated to accurately predict QoT in a live production network \cite{ofc}.

\oldrevv{In much of this previous work, strong restrictions have been placed on the variety of spectra for the the considered optical network scenario. Usually, a fixed grid and a fixed channel bandwidth is considered, varying only the modulation formats. Relaxing the limitations on the variety of transceiver configurations, especially on the center frequency when moving to a flexible grid poses a significant challenge for the feature selection of ML models for QoT estimation. Multiple parameters of each channel on the grid influence the QoT of a channel under test (CUT). The number of parameters as well as their importance on the CUT's QoT is variable and depends on the number of active channels and the position of the CUT on the considered link. It is important for the training of a generalizable and reasonably sized ML model to break down these parameters into a small-sized and fixed-length set of features that is representing the influence of the co-propagating channels onto the CUT well, in any configuration. As a solution, we have developed a divide-and-conquer approach to QoT estimation, utilizing physical models as well as ML \cite{ecoc}.}

In this work, we extend our previous proposal of a novel physical-layer-aware QoT estimation approach \cite{ecoc}. The approach uses an ML regression model and precomputed SCI values of each WDM channel as input features in order to quickly and accurately determine the total NLI for all channels. \oldrev{This approach offers a more accurate alternative to the closed-form GN model, henceforth called GN, for applications such as network planning and optimization that require fast computations.} We present an extended and comprehensive evaluation of the ML approach by including a comparison of error distributions on the test dataset. Additionally, an extensive network study is carried out in which the proposed ML-based PCE is compared against the GN in a multi-year planning scenario on different network topologies and using different RCSAs. \rev{For the studies, standard single-mode fiber networks with homogeneous span lengths are considered assuming actual load (no ASE loading).} The studies reveal that \oldrevv{the performance of GN and ML is comparable while a significant improvement in computation time with the ML PCE enables more complex network optimization. In this paper we consider three RCSA scenarios of different complexity and time granularity (Sec.~\ref{sec:rcsas}). The more complex RCSAs lead to lower underprovisioning and higher throughput. The speed advantage of using the ML PCE increases with the computational complexity of the RCSA.}

\begin{figure*}[t]
\centering
\includegraphics[width=\textwidth]{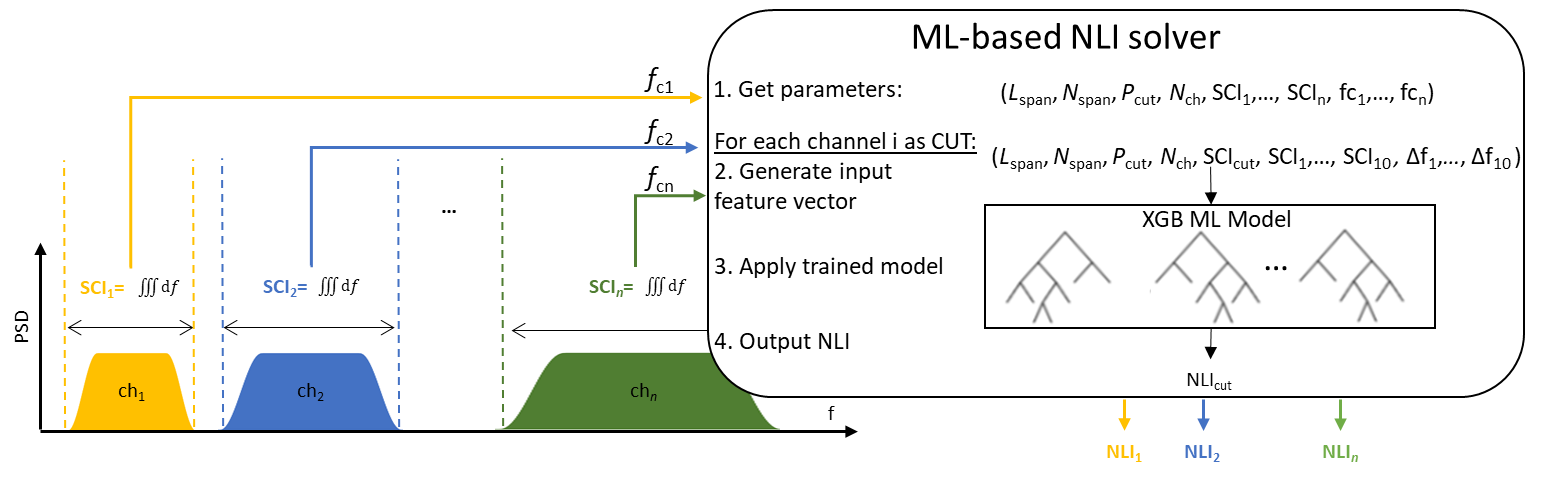}
\caption{Illustration of the proposed divide-and-conquer method for QoT estimation. SCI is computed individually for each channel \oldrev{per link} using the EGN model, and the total NLI (SCI+XCI) of a channel is determined in an ML-based NLI solver using the SCIs of neighboring channel. \oldrevv{From the precomputed SCI values and additional physical parameters, the input features are computed for each channel and finally, the NLI is determined by the XGB ML model.}}
\label{fig:NLI_Engine}
\end{figure*}
\section{Divide-and-conquer QoT Estimation} \label{sec:divide} 

We choose SNR as relevant metric for QoT. For the non-linear fiber channel, SNR is defined as \cite{egn}
\begin{equation}\label{eq:egn}
\SNR=\frac{\Ptx}{\sigma^2_\text{noise}}=\frac{\Ptx}{\sigma^2_\text{ASE} + \sigma^2_\text{SCI} + \sigma^2_\text{XCI}},    
\end{equation}
where multi-channel interference is discarded because of its minimal impact for dispersion-unmanaged standard single-mode fiber (SSMF) links \cite{egn}. The total noise computation can therefore be split into three parts. The key idea of the proposed \emph{divide-and-conquer} approach for QoT estimation is to separately choose the way of computing the different noise contributions ("divide"), thereby achieving an optimized trade-off between accuracy and speed of the QoT estimation ("conquer"). \oldrev{In a mesh network the NLI estimation is carried out on a per-link basis.}

The linear ASE noise can be calculated fast and straightforward, using well-known analytical formulas \cite[eq.(6.104)]{kumar2014fiber} assuming knowledge of the physical parameters. The SCI computation is reasonably fast, even when using the full-form integral-based EGN model. It requires integration only over a small frequency band as SCI depends only on the CUT itself. In contrast, the XCI evaluation is computationally highly complex as it needs to consider all WDM channels and thus involves several THz of integration bandwidth. In the following, we will outline the development of an QoT estimator that combines ASE computation with analytical formulas, SCI computations using the accurate full-form EGN model and an ML-based NLI solver for fast and accurate QoT estimations.

The physical motivation behind this approach is that those parameters on which the SCI of a particular WDM channel depends  define, to a large extent, the XCI that this channel inflicts onto its neighbors. Therefore the SCI of an interferer serves as a proxy for describing the interferer by its full spectrum (as we would do in the EGN model). An illustration of the principle is shown in Fig.~\ref{fig:NLI_Engine}. The proposed scheme has to solve relatively simple integrals over a small integration interval only once. \oldrevv{The SCI values, center frequencies and additional parameters such as $N_\text{span}$ and $N_\text{ch}$ are collected (1) and used to generate the input feature vectors of each channel (2). Finally the NLI is computed by the XGB ML model (3) and returned (4) for all channels.} The SCI values are stored and reused for future NLI evaluation as adding new lightpaths does not change the SCI of an existing channel. This is in contrast to conventional accurate NLI solvers, such as in \cite{egn}, that require taking into account the full spectrum to determine the NLI of each channel. 
As a first step the SCI values of all channels on the frequency grid are computed using the EGN model. The ML-based NLI solver loops through all channels, creating the input feature vector for the specific channel-under-test (CUT) and computing its NLI. The NLI computation is done using a gradient boosting model, trained and optimized as described in Section~\ref{sec:ml}. The feature vector used for the NLI computation contains the SCI value of the CUT and its 10 closest neighbors as well as their distance to the CUT. Additionally, the span length ($L_\text{span}$), number of spans ($N_\text{span}$) and launch power of the CUT are considered. Finally, combining the total NLI with the ASE noise computations according to Eq.~\ref{eq:egn}, the SNR values are returned by the QoT estimator. The evaluation of XCI, and thus total NLI, based on SCI is done in an ML model that is described in the following section.
 
\section{ML-based NLI solver}\label{sec:ml}

\begin{table}[t]
   \centering
\begin{tabular}{|l|c|}
         \hline  \textbf{Parameter} & \textbf{Range} \\
         \hline  $L_\text{span}$ [km]  & \{60, 80, 100, 120\}\\
         \hline  $N_\text{span}$  & 1 to 50, step=1\\
         \hline  Modulation  & QPSK, 16/32/64 QAM\\
         \hline \oldrev{Symbol rate [GBd]} & \oldrev{35 to 69}\\
         \hline  Data rate [Gbps]  & 100 to 600, step=50\\
         \hline
\end{tabular}
\caption{Parameter space for data generation} \label{tab:table1}
\end{table}

\subsection{Data Generation}\label{sec:data_gen}
The data for training, validation and testing of the ML model is generated using the EGN model. The EGN model allows for accurate NLI computation for a large data set with reasonable computational resources (see Sec.3.~\ref{sec:num_results}). Link and spectrum parameters are drawn randomly. \rev{The considered links are assumed to be standard single-mode fiber links of homogeneous span length uniformly drawn between 60~km and 120~km (Tab.~\ref{tab:table1}).} The QoT of each channel on the spectrum is computed. The ranges of the varied physical parameters \oldrev{are} shown in Table~\ref{tab:table1}. The spectrum is filled by adding channels with uniformly drawn parameters to the spectrum until the C-Band, divided into 12.5~GHz frequency slots, is filled \oldrev{by} 75\% up to 95\%. The power spectral density (PSD) is assumed to be equalized over all channels, with a 100G QPSK channel at 35 GBd having 0~dBm launch power. In total more than 2200 different link configurations were generated, resulting in over 230,000 data points of physical parameters of the link and the channel as well as its computed SNR. The data set was split into train/validation and test set using the conventional 70/10/20 split. In an attempt to limit the complexity of the parameter space and therefore the number of data points needed for training, note that the WDM grid layouts vary widely while the number of link layouts is limited \oldrev{by the condition of homogeneous span lengths and fixed fiber parameters (standard single-mode fiber). The work can be extended to a more general scenario by considering additional input features as} we found \cite{ofc} that links of heterogeneous span lengths and varying attenuation can be well represented by using combined parameters, such as the average of the cumulative sum of the \oldrev{effective} span lengths, as the input into an ML model. \rev{This generalization requires a substantially larger data set due to the added degrees of freedom in the parameter space, thereby also increasing the training time.}

\subsection{Choice of Model}
The ML model used for NLI computation is a gradient boosting model based on the XGBoost (XGB) \cite{xgb} library. 
In a XGB model, an ensemble of binary decision trees is constructed one by one. For a new tree a subset of all used input features is randomly selected. The tree is fitted to the prediction error of the existing ones. A prediction is made by running through all decision trees in the ensemble and adding up the attained leaf scores.
XGB models are known to perform well for regression on tabular datasets. 
We investigated the performance of an XGB model for NLI estimation in previous work and found that it outperforms other ML techniques such as neural networks while offering more control through feature importance scores and explainability. Tools such as TreeExplainer \cite{TreeExplainer} can be used to explain a gradient boosting model's output by quantifying the impact of a datapoint's input features on the model's output.
This additional interpretability helps to optimize the model but also to understand how it arrives at a prediction, building more trust in the prediction.

\subsection{Feature Selection and Model Optimization}
The input features of the model were chosen using Shapley values \cite{shap} to quantify feature importance. The most impactful physical parameters were chosen as input features by disregarding parameters with low prediction power as features, optimizing the model's performance on the validation data. The chosen input features are the SCI values of the CUT and its 10 closest WDM neighbors \oldrev{(thereby implicitly considering their launch powers)} as well as their distance to the CUT. Additionally the CUT launch power, total number of channels, span length and the number of spans are chosen as inputs. \oldrev{The number of WDM neighbors considered as well as the additional parameters were determined by the analysis of Shapley values, disregarding modulation formats, symbol rates and channel powers of neighboring channels as input.} \oldrev{In the event of less than 10 neighboring channels, the SCI and distance values of non-existent channels are set to zero.} The XGB model's hyperparameters were optimized on the training dataset using cross-validation.
The model is trained to output the NLI constant $\eta_\text{NLI}$ in dB, defined as the total NLI power normalized by the transmit power cubed. 

\begin{figure}[t]
\centering
  \input{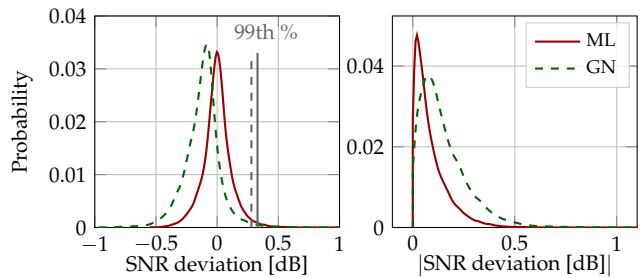}

\vspace{-.3\baselineskip}
\caption{Distribution of the estimation error in SNR (left) and the absolute estimation error (right) between ML model (red) and GN (green, dashed), in relation to EGN.}
\label{fig:distributions}
\end{figure}

\begin{figure}[t]
\centering
  \input{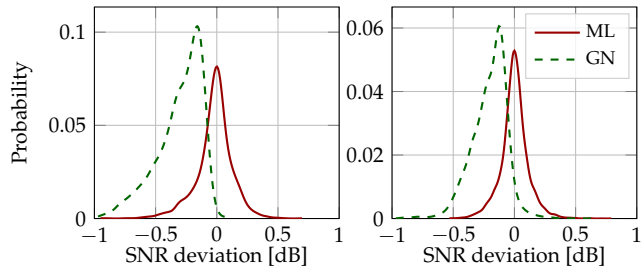}

\vspace{-.3\baselineskip}
\caption{\oldrev{Distribution of the estimation error in SNR for link lengths of less than \oldrevv{300km} (left) and for 35 GBd channels (right) between ML model (red) and GN (green, dashed), in relation to EGN.}}
\label{fig:distributions_conditioned}
\end{figure}

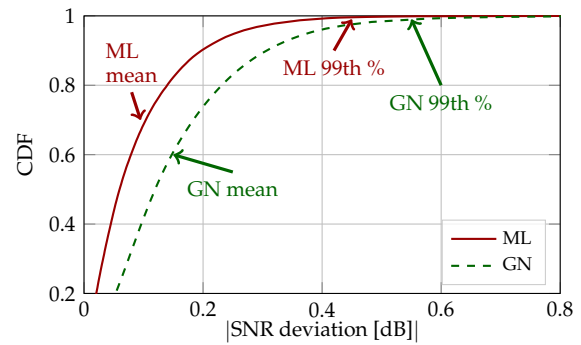
\begin{figure}[t]
\centering
\begin{tikzpicture}[font=\small]
\begin{axis}[
width=0.9\columnwidth,
height=0.6\columnwidth,
legend cell align={left},
legend style={
  fill opacity=0.8,
  draw opacity=1,
  text opacity=1,
  at={(0.97,0.03)},
  anchor=south east,
  draw=white!80!black
},
xlabel={$\left\lvert \text{SNR deviation [dB]}\right\rvert$},
ylabel={CDF},
xmin=0, xmax=0.8,
ymin=0.2, ymax=1,
xmajorgrids,
ymajorgrids,
xlabel shift = -5pt
]
\addplot [thick, red!60!black]
table {%
0.00630008352778333 0.0626339489970381
0.00734498870806988 0.0722662364244949
0.00825353478168456 0.0818985238519517
0.00917396345186461 0.0915308112794086
0.0101216566851292 0.101163098706865
0.0111880915354998 0.110795386134322
0.0121364858059048 0.120427673561779
0.0130805531608615 0.130059960989236
0.0140417212667305 0.139692248416693
0.015125532578713 0.14932453584415
0.0161985737966823 0.158956823271606
0.0172090060646255 0.168589110699063
0.0182629422123988 0.17822139812652
0.0193566272989063 0.187853685553977
0.0203321256238667 0.197485972981434
0.0213972125079991 0.207118260408891
0.0225227818166287 0.216750547836347
0.0234829496781739 0.226382835263804
0.0245216772568142 0.236015122691261
0.0256387072810256 0.245647410118718
0.0267744456212569 0.255279697546175
0.0278337193581741 0.264911984973632
0.0289834170817151 0.274544272401088
0.0301207165333413 0.284176559828545
0.0313770114544276 0.293808847256002
0.0325437090177587 0.303441134683459
0.033788462319972 0.313073422110916
0.0348317176322013 0.322705709538373
0.0360564302550088 0.332337996965829
0.0372551127998193 0.341970284393286
0.0385457704948369 0.351602571820743
0.0397585099286122 0.3612348592482
0.0410145711806891 0.370867146675657
0.0423585466845733 0.380499434103114
0.0436599307494792 0.39013172153057
0.0449602801399642 0.399764008958027
0.0462272647179853 0.409396296385484
0.0475764103093077 0.419028583812941
0.0488546888905645 0.428660871240398
0.0502384481784173 0.438293158667855
0.0516032025673745 0.447925446095311
0.0529508714190143 0.457557733522768
0.0544845958284537 0.467190020950225
0.0559292135228056 0.476822308377682
0.0574551128535532 0.486454595805139
0.0590532825979544 0.496086883232596
0.0607086965022816 0.505719170660052
0.062441157913744 0.515351458087509
0.0639898750389101 0.524983745514966
0.0656927755363412 0.534616032942423
0.0675202847672978 0.54424832036988
0.0693585457680577 0.553880607797337
0.0711249541342092 0.563512895224794
0.0731635359041984 0.57314518265225
0.0752824696125884 0.582777470079707
0.0771993301429754 0.592409757507164
0.0793425559659129 0.602042044934621
0.0814640554276949 0.611674332362078
0.0836280629908721 0.621306619789534
0.0857921480709036 0.630938907216991
0.0879521203788673 0.640571194644448
0.0903981915506149 0.650203482071905
0.0926952413589888 0.659835769499362
0.0953715698558817 0.669468056926819
0.0977629342764423 0.679100344354276
0.100560648823347 0.688732631781732
0.10332201136443 0.698364919209189
0.106045022023725 0.707997206636646
0.108954551876483 0.717629494064103
0.112044734865628 0.72726178149156
0.115155068579647 0.736894068919017
0.118382489561668 0.746526356346473
0.122086006113586 0.75615864377393
0.125736882383411 0.765790931201387
0.12926594337732 0.775423218628844
0.133176070145288 0.785055506056301
0.13762658330462 0.794687793483758
0.141436229886249 0.804320080911214
0.145691173597271 0.813952368338671
0.150240477995631 0.823584655766128
0.155060169463262 0.833216943193585
0.159963660136823 0.842849230621042
0.165127864580462 0.852481518048499
0.170924904340982 0.862113805475955
0.176488924118113 0.871746092903412
0.182862895033951 0.881378380330869
0.190104407113487 0.891010667758326
0.197845082853497 0.900642955185783
0.206819248553453 0.91027524261324
0.216426235836941 0.919907530040696
0.226788017947838 0.929539817468153
0.238048550502926 0.93917210489561
0.252343079557726 0.948804392323067
0.26864380305113 0.958436679750524
0.290243338597719 0.968068967177981
0.319732648802288 0.977701254605437
0.36286884753153 0.987333542032894
0.407579439262381 0.993112914489368
0.417884049121092 0.994076143232114
0.430184438132704 0.99503937197486
0.45195980598252 0.996002600717605
0.476646198744398 0.996965829460351
0.511409978232869 0.997929058203097
0.557510833159197 0.998892286945842
0.820859569873088 0.999855515688588
};
\addlegendentry{ML}

\addplot [thick, dashed, green!40!black]
table {%
1.90515159204097e-07 4.81609732369472e-06
0.00284987710331386 0.00963701074471313
0.00552103067173171 0.0192692053921026
0.00825569338272913 0.028901400039492
0.0110369015674774 0.0385335946868814
0.0138506464022186 0.0481657893342709
0.0166216128109404 0.0577979839816603
0.0192455605514468 0.0674301786290497
0.0219619254455239 0.0770623732764392
0.0245793973411566 0.0866945679238286
0.0271777787316285 0.096326762571218
0.0297659904509437 0.105958957218607
0.0322958063104393 0.115591151865997
0.0348384766544285 0.125223346513386
0.0373728650873684 0.134855541160776
0.0398284739418528 0.144487735808165
0.0421997212729543 0.154119930455555
0.044583581349471 0.163752125102944
0.0468735995968252 0.173384319750334
0.0491403274394671 0.183016514397723
0.051419822402611 0.192648709045112
0.0535954975901216 0.202280903692502
0.0557259702615447 0.211913098339891
0.057893557119769 0.221545292987281
0.0600463283802632 0.23117748763467
0.0622127023601777 0.24080968228206
0.064326587753416 0.250441876929449
0.0663856138558634 0.260074071576838
0.0684523666793808 0.269706266224228
0.0705249783198942 0.279338460871617
0.0725660160169195 0.288970655519007
0.0746343054864482 0.298602850166396
0.0766363640277739 0.308235044813786
0.0786824507261041 0.317867239461175
0.0806599693075274 0.327499434108564
0.0827444162682252 0.337131628755954
0.0848618113655295 0.346763823403343
0.0868657998769411 0.356396018050733
0.0888768602401964 0.366028212698122
0.0909444760415834 0.375660407345512
0.0930585195489684 0.385292601992901
0.095127191014079 0.394924796640291
0.0971716759621035 0.40455699128768
0.0993710337402511 0.414189185935069
0.101560630269866 0.423821380582459
0.103768089399995 0.433453575229848
0.105933528786079 0.443085769877238
0.108144918088087 0.452717964524627
0.11040502101384 0.462350159172017
0.112702624345975 0.471982353819406
0.115062251324161 0.481614548466795
0.117338357920064 0.491246743114185
0.11967783441464 0.500878937761574
0.122084489052932 0.510511132408964
0.124569950533932 0.520143327056353
0.127095448941098 0.529775521703743
0.129741472207174 0.539407716351132
0.132368249287984 0.549039910998521
0.135046510254082 0.558672105645911
0.137828633765489 0.5683043002933
0.140672768521346 0.57793649494069
0.143694434899665 0.587568689588079
0.146661589904696 0.597200884235469
0.149690278490256 0.606833078882858
0.152810788501082 0.616465273530247
0.156064220437178 0.626097468177637
0.159445587898716 0.635729662825026
0.162877562176462 0.645361857472416
0.16621465453027 0.654994052119805
0.169865109408754 0.664626246767195
0.173502314853136 0.674258441414584
0.177167528669079 0.683890636061974
0.180923552526078 0.693522830709363
0.184795773945957 0.703155025356752
0.188870003712463 0.712787220004142
0.193016231275168 0.722419414651531
0.197166928041323 0.732051609298921
0.201527077235166 0.74168380394631
0.206103578182566 0.7513159985937
0.210812031765915 0.760948193241089
0.215739748292176 0.770580387888478
0.220669338707079 0.780212582535868
0.225807420720226 0.789844777183257
0.231124576623927 0.799476971830647
0.236688082327301 0.809109166478036
0.242452438657754 0.818741361125426
0.248485493389126 0.828373555772815
0.2548605805738 0.838005750420205
0.261739306978017 0.847637945067594
0.268882842099796 0.857270139714983
0.276327531490555 0.866902334362373
0.284403096986992 0.876534529009762
0.292855857639008 0.886166723657152
0.302552095505945 0.895798918304541
0.312645067196542 0.905431112951931
0.323574826067516 0.91506330759932
0.335855638968017 0.924695502246709
0.349519911127073 0.934327696894099
0.364807634676978 0.943959891541488
0.383142349938344 0.953592086188878
0.406411327753961 0.963224280836267
0.437381121140426 0.972856475483657
0.483435613882932 0.982488670131046
0.577962532792199 0.992120864778435
0.595304152658258 0.993084084243174
0.616550189918662 0.994047303707913
0.640480348656144 0.995010523172652
0.670254387339522 0.995973742637391
0.706939806442172 0.99693696210213
0.756008398763463 0.997900181566869
0.821513520362643 0.998863401031608
0.960877646697085 0.999826620496347
};
\addlegendentry{GN}

\draw[<-, very thick, red!60!black] (0.095,0.7) -- (0.08,0.78) node[above,align=left] {ML\\mean};
\draw[<-, very thick, green!40!black] (0.15,0.6) -- (0.25,0.55) node[below] {GN mean};

\draw[<-, very thick, red!60!black] (0.45,0.99) -- (0.42,0.9) node[below] {ML 99th \%};
\draw[<-, very thick, green!40!black] (0.55,0.99) -- (0.6,0.8) node[below] {GN 99th \%};
\end{axis}

\end{tikzpicture}

\vspace{-.3\baselineskip}
\caption{CDF of the absolute SNR deviation between ML model (red) and GN (green, dashed), in relation to EGN.}
\label{fig:cdf}
\end{figure}

\begin{table*}[htbp!]
   \centering
\begin{tabular}{|c|c|c|c|c|c|}
\hline
\textbf{Topology} & \textbf{\# Nodes} & \textbf{\# Links} & \textbf{\# Demands} & \textbf{Avg. Node Degree} & \oldrev{\textbf{Avg. Path Length}} \\ \hline
\oldrev{Germany}~\cite{sndlib}       & 17                & 26                & 136        & 3.05   & 420 km     \\ \hline
Spain~\cite{internettopzoo}             & 16                & 27          & 120 & 3.38      & 610 km        \\ \hline
Sweden~\cite{internettopzoo}            & 25                & 29          & 300 & 2.32        & 760 km         \\ \hline
\end{tabular}%
\caption{Core network topologies considered.}
\label{tab:table-topologies}
\end{table*}

\subsection{Numerical Results on Simulation Data}\label{sec:num_results}
For an evaluation of the XGB model, the estimation error on the generated test dataset is investigated and the ML model is compared to GN. As shown in Fig.~\ref{fig:distributions} (left), the ML error distribution resembles a Gaussian distribution with a mean estimation error of 0 dB SNR and a standard deviation of 0.12 dB. The GN error, on the other hand, is biased towards underestimation of SNR with a mean error of -0.12 dB SNR, and also shows a larger standard deviation of 0.15 dB than ML. \rev{The GN model is designed to offer conservative QoT Estimations, explaining the bias towards underestimation of SNR. Nonetheless, overestimations are still observed}. The overestimation of QoT can lead to placement of lightpaths that will not transmit error free in the field. To counteract overestimation of QoT a safety margin is applied during planning. In the following we compare the 99th percentile of ML and GN model quantifying margins that need to be applied, in order to guarantee that not more than 1\% of the estimations exhibit QoT overestimation.
The 99th percentile lies at 0.28~dB SNR for GN and 0.33~dB SNR for ML. Therefore, although GN tends to underestimate SNR, overestimation of SNR is only negligibly lower than for ML. To guarantee feasibility of a lightpath, a similar safety margin has to be applied in both cases. 
\oldrev{Furthermore, while the error distribution of the ML model remains stable over different link lengths and CUT symbol rates, the error distributions of the GN model are dependent on these parameters. The estimation error of the GN model increases with lower link lengths and symbol rates. It can be seen in Fig.~\ref{fig:distributions_conditioned} that higher underestimation with a mean error of -0.2 dB SNR is observed for shorter links \rev{(less than 300~km)} as well as low symbol rates of the CUT channel (35 GBd).}
For a comparison in accuracy of GN and ML, we investigate the absolute estimation error of the respective models on the test dataset. 
The distributions of the absolute SNR errors in dB are shown on the right in Fig.~\ref{fig:distributions}. It can be seen that the ML model has a higher peak that is closer to zero and a steeper drop off than the GN. This results in  the mean absolute error of less than 0.1~dB SNR, obtained for the ML-based QoT estimator compared to 0.15~dB for the GN. The cumulative distribution functions (CDF) are shown in Fig.~\ref{fig:cdf}.
The 99th percentile of the absolute error lies below 0.45~dB SNR for the ML model, and at~0.55 dB for the GN. The ML model's computation time of just 17~\textmu s (i7-7500U CPU with 12GB RAM) is \oldrev{independent of the number of spans. This offers} a clear advantage over the GN's computation time of \oldrev{0.3ms for 1 span up to 12ms for 50 spans with an average computation time of }5ms \oldrev{on the simulation data}. ML is therefore about 300~times faster than the GN. While this disregards the time needed to compute SCI with the EGN model as input for the ML model (5-10s), for most computations in network planning applications these values do not have to be recomputed and therefore the computation can be skipped. \rev{The SCI values have to be recomputed when considering new link topologies or new candidate configurations for lightpaths.}

The full NLI (i.e., SCI+XCI) computation using the EGN model used for the data generation in Section~3.~\ref{sec:data_gen} takes on average 100 seconds. Although the millisecond time scale and the accuracy of the GN might be sufficient for a variety of applications, complex network planning tasks could require the performance and in particular the speed improvement \rev{offered by the ML model which provides computations on the nanosecond time scale ones it has been trained (around 45~min training time). The analytical models are implemented using Python and the Numpy library \cite{numpy} while ML uses the XGBoost library \cite{xgb}}. In conclusion, the proposed ML-based QoT estimator is orders of magnitude faster than the EGN model with a prediction accuracy that is comparable to previous QoT estimators \cite{ofc}.

\section{Network Planning Application}
In this section, the developed QoT estimator is applied to a network planning application. A multi-\oldrev{period} planning scenario is considered and the ML-based QoT estimator is compared to the closed-form GN model as PCE of an RCSA algorithm.

An integral part of any optical network planning study is the photonic path computation, which is used to calculate the QoT of each planned lightpath in the network. With the advent of BVTs and low-margin optical networking, modern network planning software needs to produce accurate and thereby reducing lead times on Request for Proposals (RFPs) from network operators.
Accuracy of QoT estimation plays an important role for BVT configuration selection, since multiple modulation format and data rate combinations have similar minimum receiver SNR values, and even a small variation in the NLI can lead to a reduction in the data rate configured on the BVT. \oldrev{The minimum receiver SNR thresholds were chosen according to ADVA's TeraFlex transponder \cite{adva22}.} \oldrev{This network planning application represents a capacity study carried out by simulations with certain assumptions such as exact knowledge and constancy of physical parameters. The ability to quickly run multiple studies with changing conditions such as different possible transceiver configurations and multi-fiber scenarios enables more optimized network planning, potentially leading to lower costs or higher capacity. Therefore, the computation time is important in these scenarios although computation time would not be a priority for yearly in-operation network planning. This study is meant to provide a fair comparison between the ML model and the GN model without claiming to provide the most optimal solution for meeting the demands generated by the traffic model on the given network topologies.}

In the following, we discuss the network topologies, assumptions, methodologies, and planning results for a mid-term core network planning study on three different topologies.

\begin{figure}[h!]
\centering
\includegraphics[width=\columnwidth]{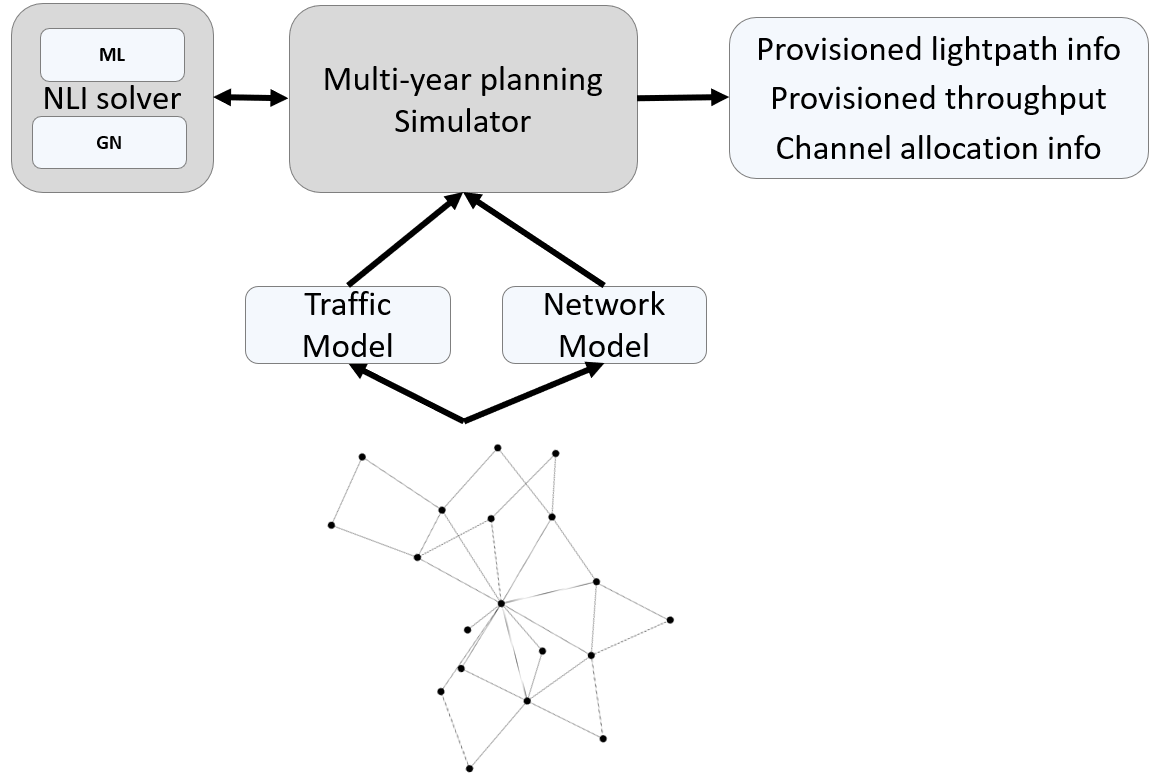}
\caption{Illustration of the planning simulator framework. The multi-year planning simulator takes inputs from the traffic model and the network model, based on the network topology. It requests QoT estimations from the GN/ML NLI-solver during each planning step and outputs the results.}
\label{fig:planning framework}
\end{figure}

\begin{figure*}[]
\captionsetup[subfigure]{justification=centering, aboveskip=-24pt}
\subfloat[\oldrev{Provisioned network throughput}]{
	\begin{minipage}[c][1\width]{
	   0.33\textwidth}
	   \centering
	   \includegraphics[width=\linewidth]{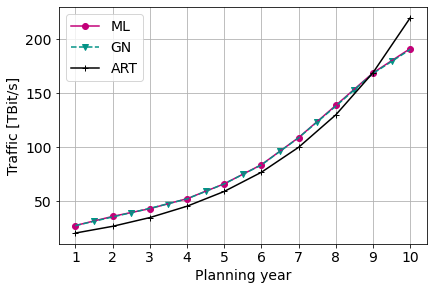}
	\end{minipage}}
 \hfill
 \hspace{-4.5\baselineskip}
\subfloat[Number of in-operation lightpaths]{
	\begin{minipage}[c][1\width]{
	   0.33\textwidth}
	   \centering
	   \includegraphics[width=\linewidth]{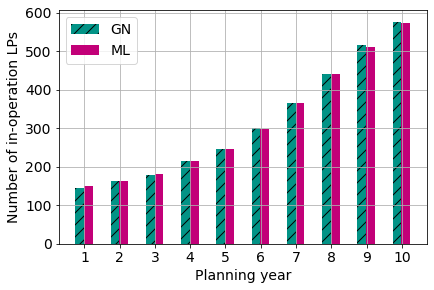}
	\end{minipage}}
 \hfill 
 \hspace{-4.5\baselineskip}
\subfloat[Underprovisioning ratio]{
	\begin{minipage}[c][1\width]{
	   0.33\textwidth}
	   \centering
	   \includegraphics[width=\linewidth]{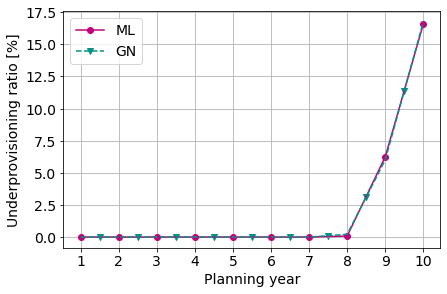}
	\end{minipage}}
	\vspace{-1.5\baselineskip}
\caption{Comparison of the ML and GN PCEs on the Spain topology using the yearly \oldrev{RCSA}.}


\label{fig:pce_comp}
\end{figure*}

\subsection{Input and Assumptions}

The impact of a QoT estimator in multi-period planning studies has been evaluated for \oldrev{multiple network planning studies} on three different network topologies. In \cite{ecoc} we investigated the small five node topology NORDUnet \cite{nordunet}. This study is extended here to larger core network topologies, namely, Nobel-Germany~(\oldrev{Germany}) \cite{sndlib}, RedIris~(Spain), and OPTOSUNET~(Sweden) \cite{internettopzoo}. The characteristics of these networks are shown in Table~\ref{tab:table-topologies}. 
While \oldrev{Germany} and Spain are sufficiently meshed core networks with an average node degree larger than 3, Sweden represents a core network topology which consists of many nodes having a node degree of 2, resulting in an average node degree of only 2.32. In such cases, the number of shortest paths available between any two given source-destination pairs is limited. \rev{A demand is considered to be the aggregated requested traffic between a source and destination node pair. Demands can be served by one or multiple lightpaths.}

To create the network model and initialize the simulation environment, several assumptions need to be made. We assume that the nodes of the topologies are reconfigurable optical add-drop multiplexer (ROADM) locations where traffic can be added, dropped, or passed through. The links are assumed to be a \oldrev{sequence} of fiber spans. However, the topologies available in \cite{sndlib,internettopzoo} do not contain fiber span information. Therefore, we assume that each link is divided into multiple SSMF spans, with each span ending in a variable gain in-line amplifier that is assumed to perform perfect attenuation compensation. The amplifiers have a noise figure of 5~dB. The complete network files are available on GitHub \cite{patriGithub}. Furthermore, we assume a flexible WDM grid scenario in the C-Band, with the availability of 400 frequency slots, each of 12.5 GHz. The ROADMs are assumed to be colorless, contentionless, and directionless, without any pass-through penalty, and only a single fiber pair is available between two ROADM locations. \rev{The filtering penalty induced by ROADMs has a minor impact of less than 1dB in these networks \cite{roadm_penalty} and is therefore ignored in the study}. The modulation format and possible data rate of each BVT is consistent with the values mentioned in Table~\ref{tab:table1}. The traffic model is based on the number of data-centers and internet exchange points in each ROADM location, and is available in \cite{patri2020planning}. This traffic model calculates the requested traffic for each demand in the initial planning period. The compounded annual growth rate of each demand follows CISCO VNI forecast of broadband technology growth in western Europe~\cite{cisco}.

\begin{figure*}[ht]
\captionsetup[subfigure]{justification=centering, aboveskip=-24pt}
\subfloat[\oldrev{Provisioned network throughput}]{
	\begin{minipage}[c][1\width]{
	   0.33\textwidth}
	   \centering
	   \includegraphics[width=\columnwidth]{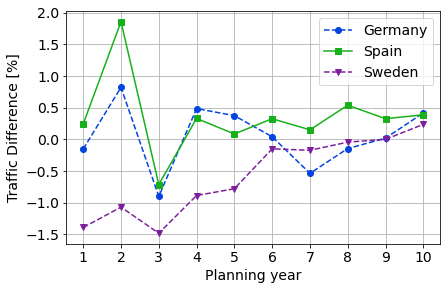}
	\end{minipage}}
 \hfill 
  \hspace{-4.5\baselineskip}
\subfloat[Number of in-operation lightpaths]{
	\begin{minipage}[c][1\width]{
	   0.33\textwidth}
	   \centering
	   \includegraphics[width=\columnwidth]{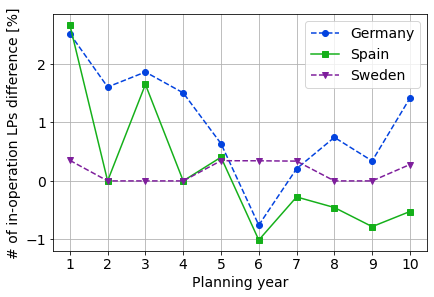}
	\end{minipage}}
 \hfill 	
  \hspace{-4.5\baselineskip}
\subfloat[Underprovisioning ratio]{
	\begin{minipage}[c][1\width]{
	   0.33\textwidth}
	   \centering
	   \includegraphics[width=\columnwidth]{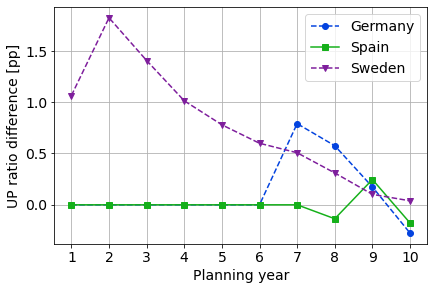}
	\end{minipage}}
	\vspace{-1.5\baselineskip}
\caption{\oldrevv{Difference between ML and GN PCEs on the three topologies.}}

\label{fig:pce_diff}
\end{figure*}

\begin{figure*}[ht!]
\vspace{-3\baselineskip}
\captionsetup[subfigure]{justification=centering, aboveskip=-24pt}
\subfloat[Germany]{
	\begin{minipage}[c][1\width]{
	   0.33\textwidth}
	   \centering
	   \includegraphics[width=\columnwidth]{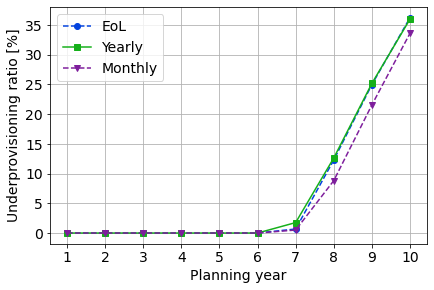}
	\end{minipage}}
 \hfill 	
  \hspace{-4.5\baselineskip}
\subfloat[Spain]{
	\begin{minipage}[c][1\width]{
	   0.33\textwidth}
	   \centering
	   \includegraphics[width=\columnwidth]{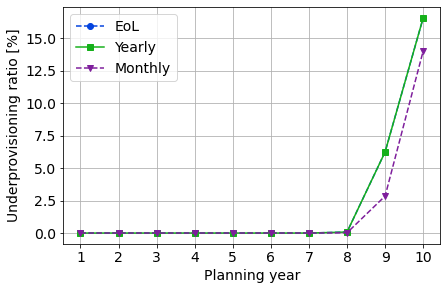}
	\end{minipage}}
 \hfill 	
  \hspace{-4.5\baselineskip}
\subfloat[Sweden]{
	\begin{minipage}[c][1\width]{
	   0.33\textwidth}
	   \centering
	   \includegraphics[width=\columnwidth]{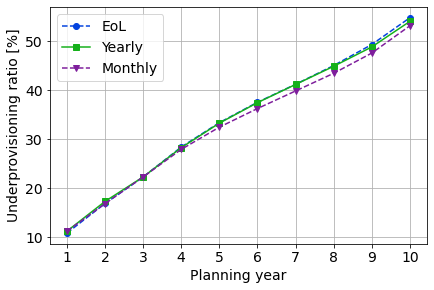}
	\end{minipage}}
	\vspace{-1.5\baselineskip}
\caption{Underprovisioning ratio for the considered RCSAs and topologies.}
\label{fig:UP}
\end{figure*}

\subsection{Planning Simulator Framework}
As shown in Fig.~\ref{fig:planning framework}, the network and traffic model serve as an input to the multi-year network planning simulator. This simulator, first introduced in \cite{patri2020planning}, creates a discrete event for each planning period and carries out the RCSA for each demand. In particular, for all the demands, a list of \textit{k}-shortest Dijkstra's simple paths is generated, where \textit{k} is the number of non-disjoint shortest paths input to the simulator. In the scope of this work, we set \textit{k=3}. \oldrev{The demands} are sorted in a decreasing first-shortest-path length order. This step is needed to reduce blocking of longer paths in the future. Further on, in our work we define a candidate lightpath as an end-to-end transparent optical signal between a BVT pair at the demand's source and destination, which could potentially carry a part of the requested traffic in a planning year. A candidate lightpath is placed only if it satisfies the spectral assignment and physical impairment constraints.

\oldrev{Three different planning studies are carried out on the topologies, varying the complexity of the RCSA scenarios. The planning is carried out for 10 years with an assumed annual growth of 30\% in traffic demands.
For each demand, a modulation format \oldrev{and path} selection multi-objective optimization function is undertaken to find the candidate lightpaths. The objective of this optimization function is to assign a modulation format and data rate combination to each BVT by minimizing the number of candidate lightpaths, then maximizing the data rate of each candidate lightpath, while meeting the traffic for each demand.}

\oldrev{In the first RCSA scenario referred as end-of-life (EoL) planning, the configurations are chosen according to end-of-life SNR, assuming the C-band is fully used on all links. For the second scenario referred as "yearly" planning, the configurations are chosen according to linear SNR. Once all the candidate lightpaths needed to satisfy the requested traffic of all demands are placed, the simulator calculates the generalised SNR, as shown in Eq.~\ref{eq:egn} and downgrades the configuration if required. \oldrevv{In case no configuration is valid, the lightpath will be removed and the demand will be considered underprovisioned.} The planning in both cases is done in a per-year granularity.
Further details, along with the constraints applied to the optimization are available~\cite{varasteh2021ondm}.
Finally, the third RCSA scenario referred as "monthly", uses the same planning tool as the "yearly" scenario but increasing the planning granularity to monthly as opposed to yearly, assuming the yearly growth is uniformly split (2.21\% per month).
}

Once the list of candidate lightpaths is generated for each demand, we allocate a central channel frequency to each candidate lightpath using a first-fit spectrum allocation algorithm in all cases. 
 The NLI part of the noise can be calculated using either the GN \oldrev{model}, or the ML \oldrev{model}.
Therefore, for each planning study, we obtain two scenarios referred as ML and GN, where ML uses the proposed divide-and-conquer XGB model (Sec.~\ref{sec:divide}), and GN uses the closed-form GN model. 

\subsection{Results and discussions}
\oldrevv{Results have been generated for ten-year planning on three core networks. We calculate the aggregate requested traffic (ART), defined as the sum of the requested data rates of all demands, the number of lightpaths placed in the network and the underprovisioning for each year.}
The simulation is first run using GN, and then using ML as PCE. Other input parameters, like the network model, traffic model, and simulation environment are kept the same for the two scenarios. \oldrev{The results show that the planning with GN and ML are close to each other with only minor differences. \rev{The advantage of ML lies in the lower computation time that significantly speeds up more complex network planning algorithms (RCSAs).}}

\subsubsection{PCE Comparison}
We compare the PCEs in provisioned throughput, number of in-operation lightpaths and underprovisioning ratio, using the yearly RCSA.
The underprovisioning ratio (UP) is hereby defined as
\begin{equation}
    \label{eq:up}
    UP =  \frac{\sum_{{\widetilde{d}}\in \widetilde{D}} \left ( DR_{\widetilde{d}} - \sum_{{lp} \in LP_{\widetilde{d}} }DR_{lp} \right )}{\sum_{{d}\in D} DR_d }.
\end{equation}
Here, $DR_{d}$ is the requested traffic of demand $d$, and $DR_{lp}$ is the data-rate of the lightpath $lp$ provisioned to carry traffic for demand $d$. $\widetilde{D}$ in Eq.~\ref{eq:up} is a subset of all the demands whose lightpaths $lp$ cannot satisfy the requested traffic, which is defined in Eq.~\ref{eq:up_d_tilde}.
\begin{equation}
    \label{eq:up_d_tilde}
    \widetilde{D} = \left \{ \widetilde{d} \in D \mid  DR_{\widetilde{d}} - \sum_{{lp} \in LP_{\widetilde{d}} }DR_{lp} > 0 \right \}
\end{equation}
Underprovisioning occurs when there are not enough contiguous free frequency slots.
Fig.~\ref{fig:pce_comp} shows that on the Spain topology the RCSA is able to satisfy all demands until planning year 8 with both PCEs. \rev{The overprovisioning seen in Fig.~\ref{fig:pce_comp}.a in this time period is due to the data rate of the lightpaths being restricted to 50 Gbit/s steps while the aggregate requested traffic does not follow this restriction.} Starting in planning year 9, underprovisioning occurs as for some demands no free spectral slot can be found on the three considered shortest paths.
While the ML PCE gives an improvement of 0.2 dB in average SNR of the deployed lightpaths, the differences in the planning are minimal. Fig.~\ref{fig:pce_comp} shows the results on the Spain topology. \rev{ML leads to about 1\% savings in the number of lightpaths in the last planning period while provisioning the same amount of throughput with the same UP through placement of higher data rate configurations in some cases.} For some lightpaths the RCSA is able to choose a higher-data rate configuration with ML than with GN reducing the number of lightpaths while provisioning the same data rate. \oldrevv{Similarly, the results on the Germany and Sweden topologies show only small differences between GN and ML with the throughput, number of LPs as well as the Underprovisioning within 2.5\% of each other for all periods as shown in Fig.~\ref{fig:pce_diff}. It can be seen that ML performs slightly worse than GN in the first periods on the Sweden topology. Due to the order of the demands a large demand with a short path can be provisioned using GN while the link is congested when using ML because earlier placed demands are provisioned with more LPs to avoid underprovisioning. This effect of the suboptimal RCSA diminishes during the later periods as all the requested data rate of all demands grows.}

\begin{figure*}[ht!]
\captionsetup[subfigure]{justification=centering, aboveskip=-24pt}
\subfloat[Germany]{
	\begin{minipage}[c][1\width]{
	   0.33\textwidth}
	   \centering
	   \includegraphics[width=\columnwidth]{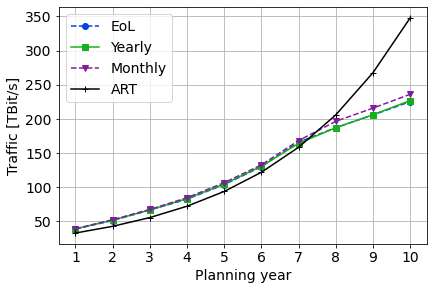}
	\end{minipage}}
 \hfill 
  \hspace{-4.5\baselineskip}
\subfloat[Spain]{
	\begin{minipage}[c][1\width]{
	   0.33\textwidth}
	   \centering
	   \includegraphics[width=\columnwidth]{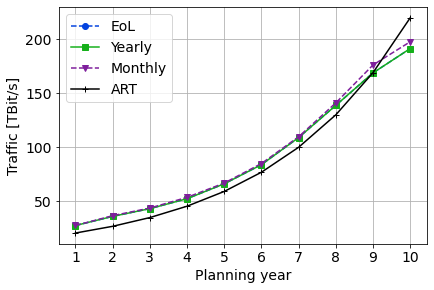}
	\end{minipage}}
 \hfill 
  \hspace{-4.5\baselineskip}
\subfloat[Sweden]{
	\begin{minipage}[c][1\width]{
	   0.33\textwidth}
	   \centering
	   \includegraphics[width=\columnwidth]{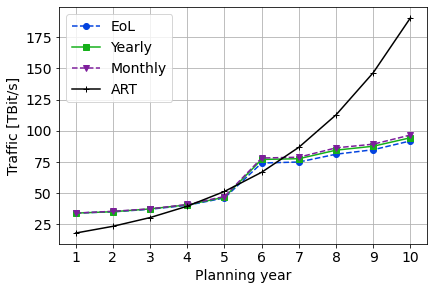}
	\end{minipage}}
	\vspace{-1.5\baselineskip}
\caption{Provisioned network throughput for the three considered RCSAs and topologies.}
\label{fig:throughput}
\end{figure*}

\begin{figure*}[ht!]
\vspace{-3\baselineskip}
\captionsetup[subfigure]{justification=centering, aboveskip=-24pt}
\subfloat[Germany]{
	\begin{minipage}[c][1\width]{
	   0.33\textwidth}
	   \centering
	   \includegraphics[width=\columnwidth]{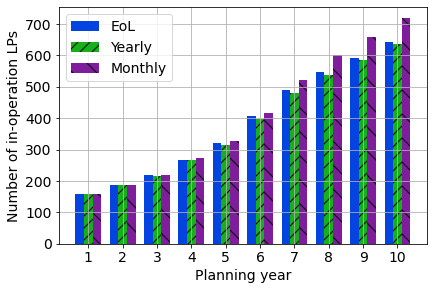}
	\end{minipage}}
 \hfill 
  \hspace{-4.5\baselineskip}
\subfloat[Spain]{
	\begin{minipage}[c][1\width]{
	   0.33\textwidth}
	   \centering
	   \includegraphics[width=\columnwidth]{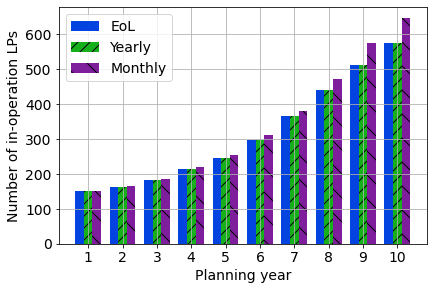}
	\end{minipage}}
 \hfill 
  \hspace{-4.5\baselineskip}
\subfloat[Sweden]{
	\begin{minipage}[c][1\width]{
	   0.33\textwidth}
	   \centering
	   \includegraphics[width=\columnwidth]{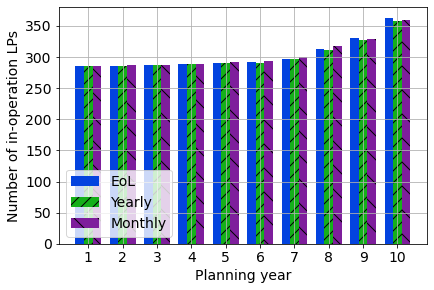}
	\end{minipage}}
	\vspace{-1.5\baselineskip}
\caption{Number of in-operation lightpaths for the three considered RCSAs and topologies.}
\label{fig:Nlp}
\end{figure*}

\begin{figure*}[ht!]
\vspace{-3\baselineskip}
\captionsetup[subfigure]{justification=centering, aboveskip=-24pt}
\subfloat[Data rate]{
	\begin{minipage}[c][1\width]{
	   0.33\textwidth}
	   \centering
	   \includegraphics[width=\columnwidth]{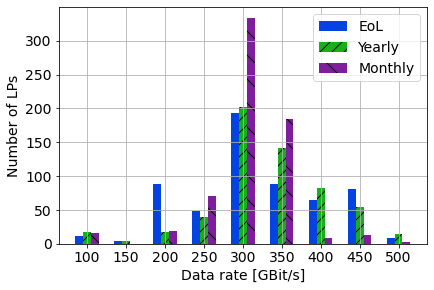}
	\end{minipage}}
 \hfill 	
  \hspace{-4.5\baselineskip}
\subfloat[Bandwidth]{
	\begin{minipage}[c][1\width]{
	   0.33\textwidth}
	   \centering
	   \includegraphics[width=\columnwidth]{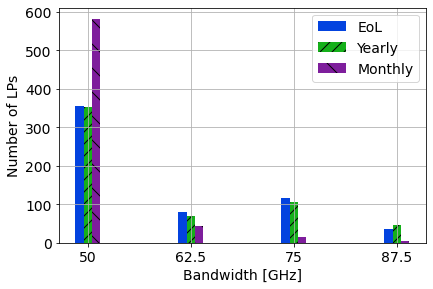}
	\end{minipage}}
 \hfill 	
  \hspace{-4.5\baselineskip}
\subfloat[Modulation format]{
	\begin{minipage}[c][1\width]{
	   0.33\textwidth}
	   \centering
	   \includegraphics[width=\columnwidth]{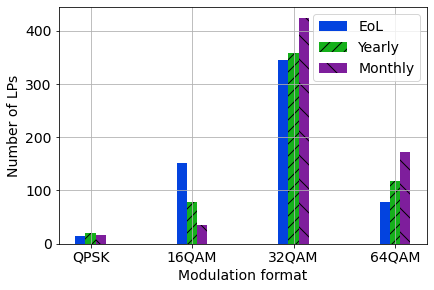}
	\end{minipage}}
	\vspace{-1.5\baselineskip}
\caption{Configuration distributions in the final planning year on the Spain topology.}
\label{fig:config_dists}
\end{figure*}

\oldrev{
\subsubsection{RCSA Comparison}\label{sec:rcsas}
\rev{Network planning algorithms of higher complexity will lead to better results and a more optimal usage of the network resources than less complex algorithms. The complexity is chosen according to computation time constraints. The PCE has a large impact on the RCSAs computation time. In the following, we compare the three RCSA scenarios using the ML PCE on three different topologies and compare the computation time to using the GN PCE.} As shown in Tab.~\ref{tab:table-topologies} the three topologies differ in average node degree and path length. While all RCSA scenarios can meet the demands for the Germany topology up to planning year 7 with an ART of over 150 TBit/s, \rev{for Sweden, underprovisioning of demands (Fig.~\ref{fig:UP}.c)  can be observed from the first year with an ART of less than 24 TBit/s as the routing options are limited due to the low average node degree leading to bottleneck links being part of the shortest routes of many demands. Additionally, the long average path lengths leads to lower data rate configurations being chosen by the RCSAs and the spectrum filling up fast.} On the Sweden topology the provisioned throughput increases only slightly for most planning years with a large step observed in planning year 6. In this planning year lightpaths provisioned for initially small demands (< 25 GBit/s) are upgraded to account for the increasing requested data rates leading to a significant increase in provisioned throughput.}

\oldrev{Fig.~\ref{fig:throughput} shows that the same amount of network traffic is deployed by all three RCSAs on the Germany and Spain topology as long as all demands can be met and no underprovisioning occurs. It can be seen in Fig.~\ref{fig:Nlp} that the end-of-life RCSA needs a higher number of lightpaths to reach the same throughput compared to the other two scenarios in these planning years, with the yearly RCSA deploying the lowest number of lightpaths. In the planning years where underprovisioning occurs the monthly RCSA has the lowest UP on all topologies as shown in Fig.~\ref{fig:UP}, outperforming the other scenarios.}
\oldrev{This is due to the fact that smaller increase in demands per planning period in the monthly planning scenario favors lower bandwidth candidate lightpath selection, leading to the possibility of higher modulation formats being used. This can be confirmed by the configuration distributions of the final planning year on the Spain topology, as shown in Fig.~\ref{fig:config_dists}. It is further observed that high-bandwidth lightpaths need to be downgraded in modulation format more often during later planning years due to higher NLI. Therefore, the monthly planning scenario leads to a higher spectral efficiency in the final planning year using smaller bandwidth lightpaths with higher modulation formats. Of the considered RCSAs, the monthly RCSA minimizes UP but the yearly RCSA is the preferable solution when enough spectrum is available as it minimizes the number of deployed lightpaths.
}

\oldrev{
\subsubsection{Computation time}
The considered RCSAs increase in computational complexity from end-of-life to monthly as can be seen in Tab.~\ref{tab:times}. The PCE has a considerable impact on the computation time of the RCSA. It can be seen that while ML leads to higher computation times of the end-of-life RCSA due to the pre-calculation of SCI values, it enables an improvement of over 50\% for the yearly RCSA and up to 70\% for the monthly RCSA. The improvement in computation time scales with the complexity of the RCSA and the number of required QoT estimations. The highest improvement of 70\% can therefore be observed using the monthly RCSA on the Sweden topology.
}

\begin{table*}[t]
   \centering
\begin{tabular}{|c|c|c|c|c|}
         \hline \textbf{Topology} & \textbf{PCE} & \textbf{end-of-life RCSA} & \textbf{yearly RCSA} & \textbf{monthly RCSA} \\
         \hline  \multirow{2}{*}{\textbf{Germany}} & GN & 395 & 2254 & 10962  \\\cline{2-5}
                                             & ML  & 486 & 1042 & 3498 \\
        \hline  \multirow{2}{*}{\textbf{Spain}} & GN & 372 & 1947 & 10829 \\\cline{2-5}
                                             & ML  & 460 & 944 & 3270 \\
        \hline  \multirow{2}{*}{\textbf{Sweden}} & GN & 968 & 4926 & 14326 \\\cline{2-5}
                                             & ML  & 1023 & 2207 & 4278 \\
         \hline
\end{tabular}
\caption{Computation times in seconds for the three considered RCSAs and topologies.} \label{tab:times}
\end{table*}

\section{Conclusion}
We developed a QoT estimation method utilizing an XGB model to quantitatively predict NLI. The model \oldrevv{is trained for flexible-grid networks and} takes SCI values computed by the EGN model as input. The ML-based NLI solver achieves high accuracy on simulation data with a mean absolute estimation error of less than 0.1 dB SNR. It is shown to outperform the traditional approach of using a closed-form GN model as NLI solver, avoiding the bias towards underestimation of SNR. We conducted an extensive multi-period network planning study on three different network topologies, comparing the use of ML and GN as PCE to investigate the effect of ML's improved accuracy and faster computation time in a network planning context. Using the chosen SCI and distance features the ML model is able to generalize well as the performance of the ML and GN PCEs is comparable on all three topologies. 
\rev{We show that using an ML-based PCE in network planning can lead to significant improvements in computation time of up to 70\%. In the presence of computational constraints, more complex RCSAs can be considered when using the ML PCE resulting in more efficient usage of the available capacity in a network and serving as part of a cost-efficient solution for the quickly rising demands in network throughput.} 

\section*{Disclosures.}
The authors declare no conflicts of interest.

\section*{Acknowledgements}
This work has been partially funded in the framework of the CELTIC-NEXT project AI-NET-PROTECT (Project ID C2019/3-4) by the German Federal Ministry of Education and Research (\textbf{\#16KIS1279K}). Carmen Mas-Machuca acknowledges the support by the Federal Ministry of Education and Research of Germany (BMBF) in the programme of "Souverän. Digital. Vernetzt." joint project 6G-life (\textbf{\#16KISK002}). 

\bibliography{references}

\end{document}